\documentclass[a4paper,showkeys,amsmath,amssymb,pra,twocolumn]{revtex4-1}
\usepackage[utf8]{inputenc}
\usepackage{amsmath,amssymb}
\usepackage{mathtools}
\usepackage{multirow}
\usepackage{graphicx}
\usepackage{subfig}
\usepackage{bbm}
\usepackage{hyperref}
\usepackage[qm]{qcircuit}

\newcommand{\ket}[1]{| #1 \rangle}
\newcommand{\bra}[1]{\langle #1 |}

\newcommand{\braket}[2]{\bra{#1}{#2}\rangle}

\newcommand{\tr}{\mathrm{Tr}}

\newcommand{\Id}{\mathbbm{1}}

\newcommand{\ii}{\mathrm{i}}

\newcommand{\HH}{\mathcal{H}}

\newcommand{\ie}{\emph{i.e.\/}}

\begin{document}
\author{{\L}ukasz Pawela}
\email{lpawela@iitis.pl}
\affiliation{Institute of Theoretical and Applied Informatics, Polish Academy
of Sciences, Ba{\l}tycka 5, 44-100 Gliwice, Poland}

\title{Quantum games on evolving random networks}
\date{30/12/2015}

\begin{abstract}
We study the advantages of quantum strategies in evolutionary social dilemmas 
on evolving random networks. We focus our study on the two-player games: 
prisoner's dilemma, snowdrift and stag-hunt games. The obtained result show the 
benefits of quantum strategies for the prisoner's dilemma game. For the other 
two games, we obtain regions of parameters where the quantum strategies 
dominate, as well as regions where the classical strategies coexist.
\end{abstract}

\maketitle

\section{Introduction}
Game theory is a widely studied branch of science with broad applications in a
plethora of fields. These range from biology to social sciences and economics.
It has been especially useful in the study of social dilemmas, \ie situations
where the benefit of the many should be put in front of the benefit of the
individual. One of the most frequently studied approaches in this context is the
evolutionary game theory~\cite{nowak1992evolutionary}. The field of evolutionary
games has since evolved and now studies not only games on regular grids, but
also on complex graphs~\cite{szabo2007evolutionary}. Recently, there are studies
focused on studying social dilemmas on evolving random
networks~\cite{albert2000topology,szolnoki2009resolving}.

In quantum game
theory~\cite{eisert1999quantum,meyer1999quantum,flitney2002introduction,piotrowski2003invitation}
 we allow the agents to use quantum strategies alongside classical ones. As 
this is a far larger set of possible players' moves, it offers the possibility 
of much more diverse behavior. The most outstanding example of this, is the 
fact that if only one player is aware of the quantum nature of the game, he/she 
will never lose in some types of games. If both players are aware of the nature 
of the game, one of them might still cheat by appending additional qubits to 
the system~\cite{miszczak2012qubit}. When we take decoherence into account, the 
game behavior changes. In particular, the well known Nash equilibrium of a game 
can shift to a different strategy~\cite{gawron2014decoherence}. Furthermore, 
quantum game theory allows us to solve dilemmas present in classical game 
theory, like the prisoner's dilemma~\cite{flitney2003advantage}. On top of 
this, there 
also exists a quantum version of the Parrondo's 
paradox~\cite{flitney2002quantum,pawela2013cooperative}. Finally, there quantum 
pseudo-telepathy games. In these games, players utilizing quantum strategies 
and quantum entanglement, may seem to an outside observer as communicating 
telepathically~\cite{brassard2005quantum,gawron2008noise,pawela2013enhancing}.

The combination of the fields of quantum game theory and evolutionary games has
led to numerous
results~\cite{li2012quantum,li2012evolution,miszczak2014general,pawela2013quantum}.
There exist cases where the quantum strategies dominate the entire network. In 
this work we aim to study the behavior of three quantum games on evolving 
random networks: prisoner's dilemma, snowdrift and stag-hunt games. The 
transition between these games will be achieved by manipulating the parameters 
of the game. Games on evolving networks have been studied in the 
classical~\cite{szabo2007evolutionary,szolnoki2009resolving} as well as quantum 
settings~\cite{li2013coevolution}. The evolution of of the network can be seen 
as aging of the agents.

This paper is organized as follows. In Sec.~\ref{sec:quantum-games} we
introduce quantum games. In Sec.~\ref{sec:networks} we discuss the model of
networks used in this work. Sec.~\ref{sec:results} contains the results along
with discussions. Finally, in Sec.~\ref{sec:final} we draw the final
conclusions.

\section{Quantum games}\label{sec:quantum-games}
We call a game a \emph{quantum game} if the players participating are allowed
to use quantum strategies. By quantum strategies we understand moves that have
no justification in classical game theory, but have a good interpretation in
the realm of quantum mechanics. We will focus on two-player games. Henceforth
we will call the players Alice and Bob.

\subsection{General concepts}
Formally, a two-player quantum game is a tuple $\Gamma = (\HH, \rho,
S_\mathrm{A}, S_\mathrm{B}, P_\mathrm{A}, P_\mathrm{B})$. Here $\HH$ is a
Hilbert space of the system used in the quantum game, $\rho$ is the system's
initial state. Note that $\rho$ is a positive operator with unit trace, \ie
$\rho \geq 0$ and $\tr \rho = 1$. Allowed Alice's and Bob's strategies are
given by the sets $S_\mathrm{A}$ and  $S_\mathrm{B}$ respectively. Their payoff
functions are given by $P_\mathrm{A}$ and $P_\mathrm{B}$. They are funtions
mapping players strategies to numerical values. In general the strategies
$s_\mathrm{A} \in S_\mathrm{A}$ and $s_\mathrm{B} \in S_\mathrm{B}$ can be any
quantum operations. A definition of a quantum game may contain additional rules
like the ordering of players or the number of times they are allowed to make a
move.

By analogy to the classical game theory we may define the following quantities 
in quantum game theory. We will call a strategy $s_\mathrm{A}$ the 
\emph{dominant strategy} of Alice if $P(s_\mathrm{A}, s'_\mathrm{B}) \geq 
P(s'_\mathrm{A}, s'_\mathrm{B})$ for all $s_\mathrm{A} \in S_\mathrm{A}$, 
$s'_\mathrm{B} \in S_\mathrm{B}$. Fallowing this pattern we may define a 
dominant strategy for Bob. A pair of strategies $(s_\mathrm{A}, s_\mathrm{B})$ 
is an equilibrium in dominant strategies if and only if 
$s_\mathrm{A}$ and $s_\mathrm{B}$ are Alice's and Bob's dominant strategies. A 
pair of strategies is \emph{Pareto optimal} if it not possible to increase one 
player's payoff without decreasing the other player's payoff. Finally, we 
define a Nash equilibrium as a set of strategies, such that no player can do 
better by unilaterally changing their strategy. This will be further discussed 
when we introduce the quantum prisoner's dilemma game.

\subsection{Quantizing the prisoner's dilemma, stag hunt and snowdrift games}
\begin{figure}[htp!]
\[
\Qcircuit @C=2.8em @R0.7em {
\lstick{\ket{0}} & \multigate{1}{J} & \gate{U_\mathrm{A}} & 
\multigate{1}{J^\dagger} & \qw\\
\lstick{\ket{0}} & \ghost{J} & \gate{U_\mathrm{B}} & \ghost{J^\dagger} & \qw
}
\]
\caption{Quantum circuit depicting a two-player quantum game. Here $J$ is the 
entangling operator and $U_\mathrm{A}$ and $U_\mathrm{B}$ denote Alice's and 
Bob's strategy respectively.} \label{fig:game}
\end{figure}
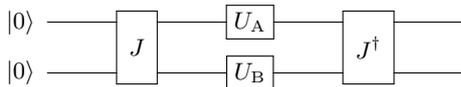

The setup in the quantum case is as follows. Each player is given by a referee
a single qubit and may only operate on it locally. Hence we have $s_\mathrm{A},
s_\mathrm{B} \in SU(2)$, where $SU(2)$ is the set of unitary $2 \times 2$
matrices with unit determinant. Initially, the qubits are entangled:
\begin{equation}
\ket{\phi} = J \ket{00},
\end{equation}
where $J$ is the entangling operator~\cite{benjamin2001multiplayer}:
\begin{equation}
J = \frac{1}{\sqrt{2}}(\Id \otimes \Id + \ii \sigma_x \otimes \sigma_x).
\end{equation}
Here $\sigma_x$ is the Pauli matrix:
\begin{equation}
\sigma_x = 
\begin{pmatrix}
0 & 1 \\
1 & 0
\end{pmatrix}.
\end{equation}
Next, the players apply their respective strategies $U_\mathrm{A}$ and 
$U_\mathrm{B}$ and the untangling operator $J^\dagger$ is applied by the 
referee. Here $J^\dagger$ denotes the Hermitian conjugate of $J$. The final 
state of the system is:
\begin{equation}
\ket{\psi} = J^\dagger (U_\mathrm{A} \otimes U_\mathrm{B}) J \ket{\phi}.
\end{equation}
This is shown as a quantum circuit in Fig.~\ref{fig:game}.

The payoff matrix of a two player game with cooperators $C$ and defectors $D$ is
shown in Tab.~\ref{tab:payoffs}. In the table $R$ is the reward, $P$ is the
punishment for mutual defection, $S$ is know as the sucker's payoff and finally
the parameter $T$ is the defector temptation. In our analysis we set $R=1$ and
$P=0$. The remaining two parameters range is $S \in [-1, 1]$ and $T \in [0, 2]$.
When $T > R > P > S$ we get a social dilemma - the prisoner's dilemma. Note that
on the one hand, in this case the strategy profile $(C, C)$ is Pareto optimal,
but on the other hand the profile $(D, D)$ is a Nash equilibrium. Hence, we have
the dilemma. Next, when $T > R > S > P$ we get the snowdrift game. Finally, when
$R>T>P>S$ we get the stag-hunt game.
\begin{table}[!h]
\begin{tabular}{c|cc}
& Bob: $C$ & Bob: $D$ \\
\hline
Alice: $C$ & ($R$, $R$) & ($S$, $T$) \\
Alice: $D$ & ($T$, $S$) & ($P$, $P$)
\end{tabular}
\caption{Payoff matrix of a two-player game with cooperators $C$ and defectors 
$D$.} 
\end{table}\label{tab:payoffs}

In the quantum case, the payoff of Alice is determined by:
\begin{equation}
P_\mathrm{A} = R |\braket{\psi}{00}|^2 + S |\braket{\psi}{01}|^2 + T 
|\braket{\psi}{10}|^2 + P |\braket{\psi}{11}|^2.
\end{equation}

Aside from the two classical strategies, we introduce two quantum strategies.
The first one is the Hadamard strategy, $H$. It introduces a \emph{miracle
move}~\cite{eisert1999quantum} in the prisoners dilemma game, that is it allows
the player to always win against the other player's classical strategy. The
second strategy, denoted by $Q$ will introduce a new Nash equilibrium in the
prisoners dilemma game, for the strategy profile $(Q, Q)$.

We associate the player's strategies with the following unitary matrices:
\begin{equation}
\begin{array}{cc}

C = \begin{pmatrix}
1 & 0 \\
0 & 1
\end{pmatrix},

&

D = \begin{pmatrix}
0 & 1 \\
1 & 0
\end{pmatrix},

\\

&

\\

H = \begin{pmatrix}
1 & 1 \\
1 & -1
\end{pmatrix},

&

Q = \begin{pmatrix}
\ii & 0 \\
0 & -\ii
\end{pmatrix}.

\end{array}
\end{equation}
\section{Simulation setup}\label{sec:networks}
We set the population size to $2500$ agents located at the nodes of an
Erd{\"o}s--R{\'e}nyi graph. The initial number of edges is set to $10000$. Each 
agent is assigned an initial strategy at random. We study the following initial 
assignments:
\begin{enumerate}
\item The classical strategies, $S_1 = \{C, D\}$. Initial these strategies are 
assigned with probability $50\%$ each.

\item Classical strategies and the miracle move, $S_2 = \{C, D, H\}$. The 
initial probabilities of assignment of $C$, $D$ and $H$ are $49\%$, $49\%$ and 
$2\%$ respectively.

\item Classical strategies and the quantum Nash equilibrium, $S_3 = \{C, D, 
Q\}$. The initial probabilities of assignment of $C$, $D$ and $Q$ are $49\%$, 
$49\%$ and $2\%$ respectively.

\item All four strategies, $S_3 = \{C, D, H, Q\}$. The initial probabilities of
assignment of $C$, $D$, $H$ and $Q$ are $49\%$, $49\%$, $1\%$ and $1\%$
respectively.
\end{enumerate}

The evolution of the network is performed via Monte Carlo simulation. First, we 
select a random agent $x$ and his random neighbor, $y$. Next, each of them 
acquires their payoff, $p_x$ and $p_y$ respectively by playing with all of 
their neighbors. Finally, if $p_x > p_y$, agent $y$ may adopt the strategy of 
agent $x$ with probability:
\begin{equation}
W = \frac{p_x-p_y}{\alpha \max(k_x, k_y)},
\end{equation}
where $k_x$ and $k_y$ are degrees of nodes $x$ and $y$ respectively and $\alpha$
is a constant dependent on the game. We have $\alpha = T-S$ for the prisoner's
dilemma, $\alpha=T-P$ for the snowdrift and $\alpha=R-S$ for the stag-hunt game.
When an agent $x$ adopts a new strategy, we remove all edges connecting him/her
to other agents, except for the one connecting to the donor agent. As this
scheme will quickly lead to the creation of disjoint graphs, we allow the agents
to form new links at the end of each Monte Carlo step. Each agent is allowed to
connect to a uniformly randomly chosen node he/she is not connected to. To avoid
creating very big graphs, we set a limit on an agent's degree, $k_\mathrm{max}$.
We set $k_\mathrm{max}=500$. If an agent has a degree greater or equal to
$k_\mathrm{max}$, we remove all outgoing edges, except for one, randomly chosen.
The agent keeps his/hers strategy. This process simulates the aging and dying of
agents. On average, every agent is chosen once per Monte Carlo step. We set the
number of steps to $10^4$. We obtain the fractions of strategies by averaging
the last $10^3$ steps.
\section{Results and discussion}\label{sec:results}
\begin{figure}[!h]
\subfloat[$\rho_C$]{\includegraphics{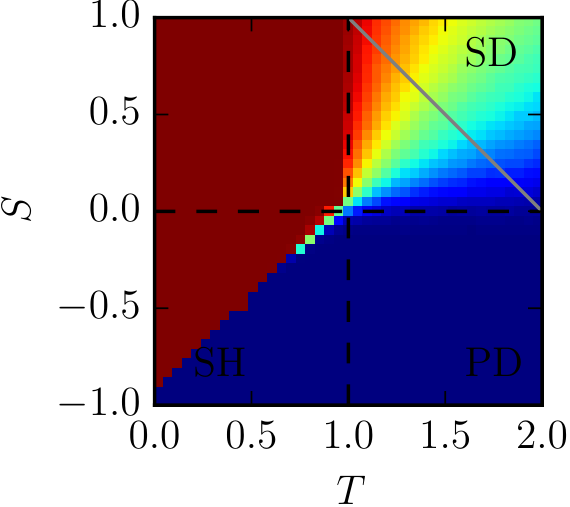}\label{fig:classical-a}}\\
\subfloat[$\rho_D$]{\includegraphics{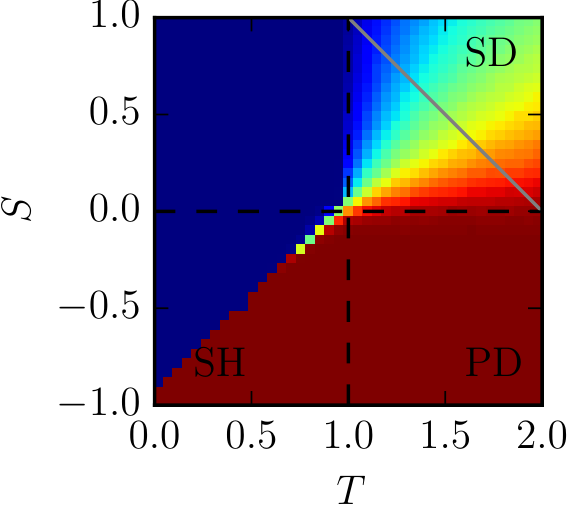}\label{fig:classical-b}}\\
\centering\includegraphics{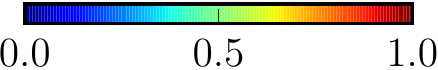}

\caption{Results for the classical strategies. Panel
\protect\subref{fig:classical-a} shows the fraction of $C$ strategy, $\rho_C$,
and panel \protect\subref{fig:classical-b} shows the fraction of strategy $D$,
$\rho_D$. Dashed black lines mark the boundaries between the different game
types. The labels correspond to prisoner's dilemma (PD), snowdrift (SD) and
stag-hunt games (SH). The solid gray lines show the regions 
which were examined in detail.}\label{fig:classical}
\end{figure}

\begin{figure}[!h]
\centering\includegraphics{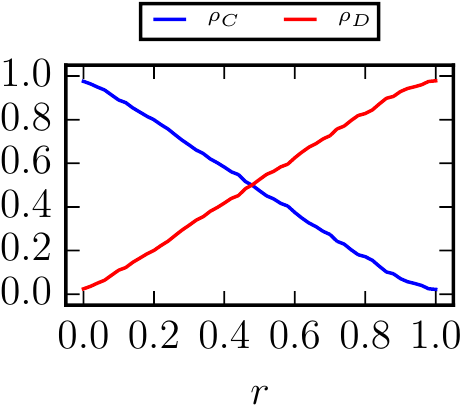}

\caption{Behavior of the fraction of strategy $C$, $\rho_C$ and the fraction of 
strategy $D$, $\rho_D$ along the line shown in Fig.~\ref{fig:classical}. The 
parameter $r$ gives the values of $S=1-r$ and 
$T=1+r$.}\label{fig:classical-line}
\end{figure}

\begin{figure}[!h]
\subfloat[$S=0.4$, 
$T=1.6$]{\includegraphics[width=0.25\textwidth]{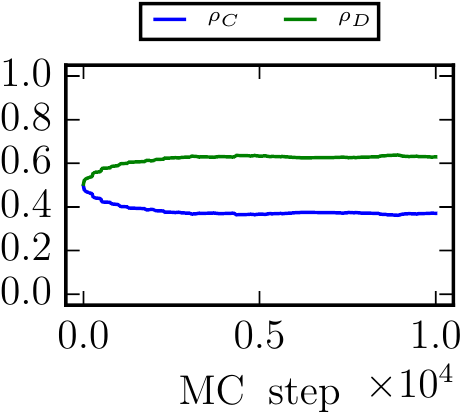}}
\subfloat[$S=0.5$, 
$T=1.5$]{\includegraphics[width=0.25\textwidth]{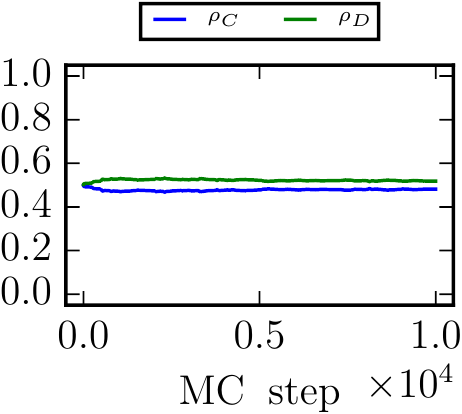}}\\
\subfloat[$S=0.6$, 
$T=1.4$]{\includegraphics[width=0.25\textwidth]{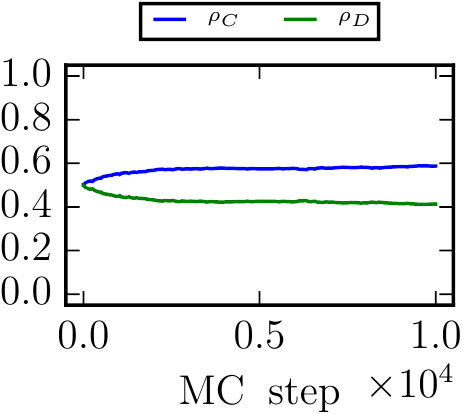}}

\caption{Full Monte Carlo history of the fractions of strategies $C$ denoted  
$\rho_C$ and $D$ denoted $\rho_D$.}\label{fig:classical-line-hist}
\end{figure}

First, we present the results for the classical strategies alone. These are
shown in Fig.~\ref{fig:classical}. These are with good agreement with previously
found results~\cite{szolnoki2009resolving}. Note that in the snowdrift game we
find the region, where the fractions of strategies transfer smoothly from the
dominance of the $C$ to the dominance of the $D$ strategy. We study one line
across this region, shown as the solid gray line in Fig.~\ref{fig:classical}.
Instead of averaging, we show only the fractions for the last iterations of the
Monte Carlo process. This is shown in Fig.~\ref{fig:classical-line}. In
Fig.~\ref{fig:classical-line-hist} we show full Monte Carlo history for a few
selected points along the line in Fig.~\ref{fig:classical-line}. This shows that
the strategies achieve their equilibrium fractions quickly, with only minor
changes after the first $2 \cdot 10^3$ steps. The parameter $r$ in
Fig.~\ref{fig:classical-line} gives the parameters $S$ and $T$ as:
\begin{equation}
\begin{split}
S &= 1 - r \\
T &= 1 + r
\end{split}.
\end{equation}
Note that al the figures show a very smooth transition between the $C$ and $D$ 
dominance case.


\begin{figure}[!h]
\subfloat[$\rho_C$]{\includegraphics{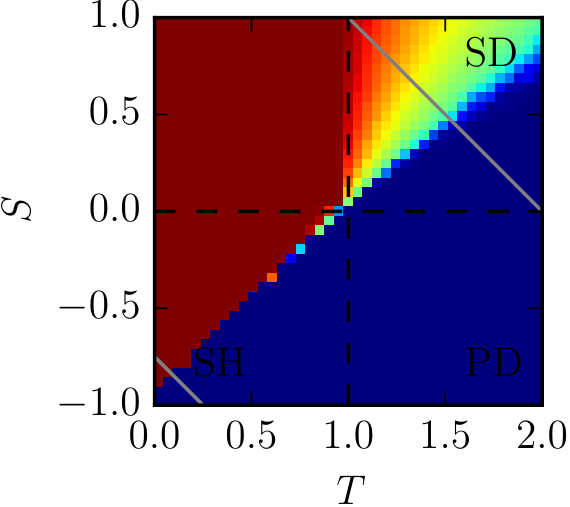}\label{fig:hadamard-a}}\\
\subfloat[$\rho_D$]{\includegraphics{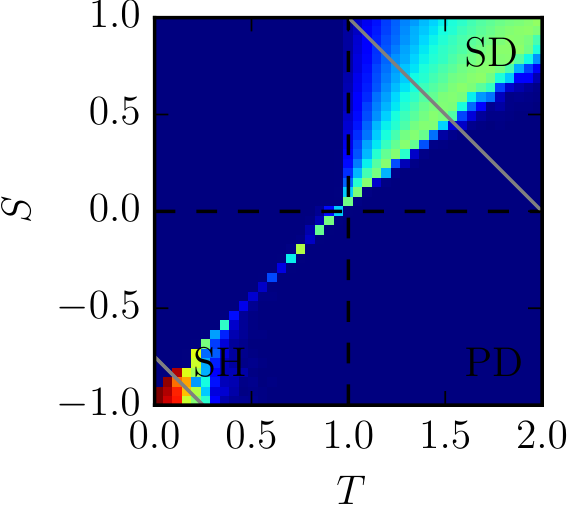}\label{fig:hadamard-b}}\\
\subfloat[$\rho_H$]{\includegraphics{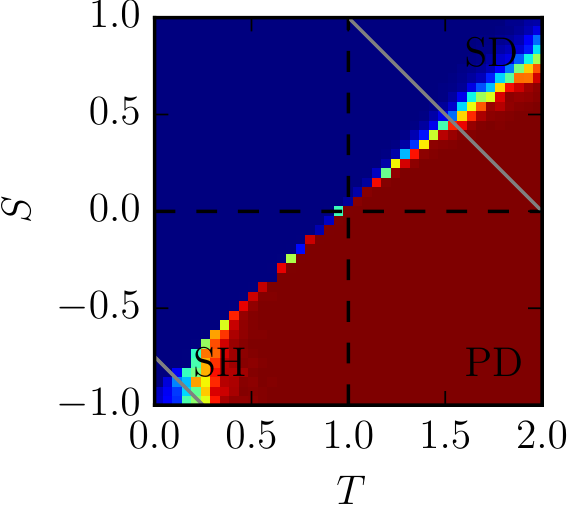}\label{fig:hadamard-c}}\\
\centering\includegraphics{colorbar}

\caption{Results for the classical strategies. Panel
\protect\subref{fig:hadamard-a} shows the fraction of $C$ strategy, $\rho_C$,
panel \protect\subref{fig:hadamard-b} shows the fraction of strategy $D$,
$\rho_D$ and panel \protect\subref{fig:hadamard-c} shows the fraction of the
strategy $H$, $\rho_H$. Dashed black lines mark the boundaries between the
different game types. The labels correspond to prisoner's dilemma (PD),
snowdrift (SD) and stag-hunt games (SH). The solid gray lines show the regions 
which were examined in detail.}\label{fig:hadamard}
\end{figure}

\begin{figure}[!h]
\centering\includegraphics{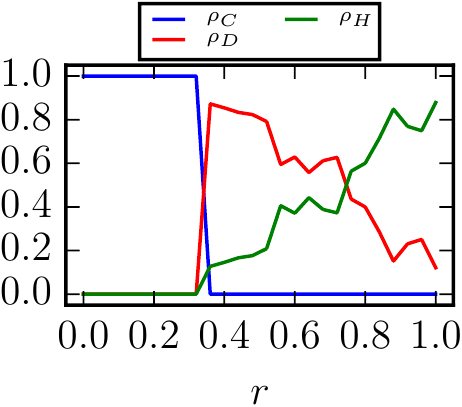}

\caption{Behavior of the fraction of strategy $C$, $\rho_C$, the fraction of
$D$, $\rho_D$ and the fraction of strategy $H$, $\rho_H$ along the line shown in
the lower left corner of Fig.~\ref{fig:hadamard}. The parameter $r$ gives the
values of $S=-\frac34-\frac14 r$ and $T=\frac14 r$.}\label{fig:hadamard-line-sh}
\end{figure}

\begin{figure}[!h]
\subfloat[$S=-0.83$, 
$T=0.08$]{\includegraphics[width=0.25\textwidth]{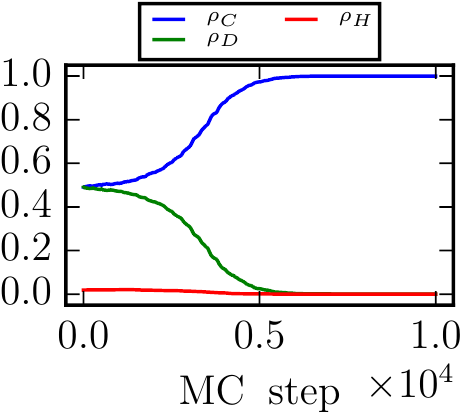}}
\subfloat[$S=-0.84$, 
$T=0.09$]{\includegraphics[width=0.25\textwidth]{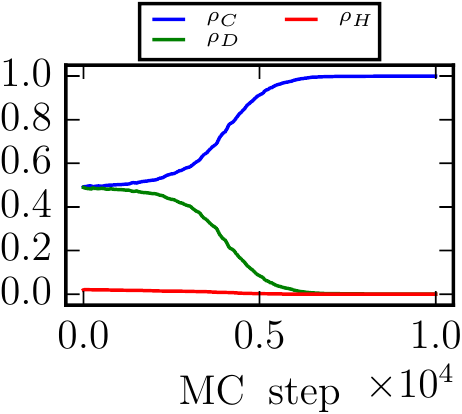}}\\
\subfloat[$S=-0.94$, 
$T=0.19$]{\includegraphics[width=0.25\textwidth]{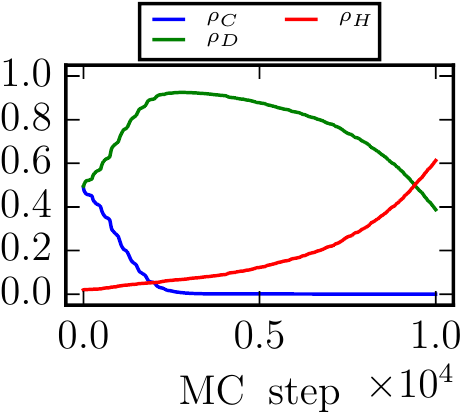}}

\caption{Full Monte Carlo history of the fractions of strategies $C$, denoted
$\rho_C$ , $D$ denoted $\rho_D$ and $H$, denoted 
$\rho_H$.}\label{fig:hadamard-line-hist-sh}
\end{figure}

Next, we study the case where we introduce the miracle move strategy, $H$. The
fractions of strategies are shown in Fig.~\ref{fig:hadamard}.In this case we
obtain different behavior compared to the classical case. First of all, the
strategy $D$ does not dominate in the prisoner's dilemma case. This region of
parameters $T$ and $S$ is now dominated by the quantum strategy $H$. In the
stag-hunt game, a transition between the dominance of $D$ and $H$ emerges. In
some regions the transition is smooth, while other regions show a sharp
transition. This is studied in the same manner as described in the previous
paragraph. We show this results in Figs.~\ref{fig:hadamard-line-sh} and
\ref{fig:hadamard-line-hist-sh}. We should note here, that we observe two 
behaviors here which depend on the parameter's values. In the first case the 
system quickly converges to total dominance of the $C$ strategy. In the second 
case, first the fraction of $D$ strategy rises, and next starts to drop in 
favor of the miracle move $H$.

\begin{figure}[!h]
\centering\includegraphics{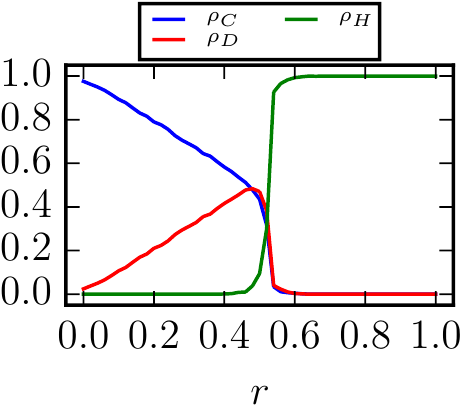}

\caption{Behavior of the fraction of strategy $C$, denoted $\rho_C$, the
fraction of $D$, denoted $\rho_D$ and the fraction of strategy $H$, denoted
$\rho_H$ along the line shown in the upper right corner of
Fig.~\ref{fig:hadamard}. The parameter $r$ gives the values of $S=1-r$ and
$T=1+r$.}\label{fig:hadamard-line}
\end{figure}

\begin{figure}[!h]
\subfloat[$S=0.6$, 
$T=1.4$]{\includegraphics[width=0.25\textwidth]{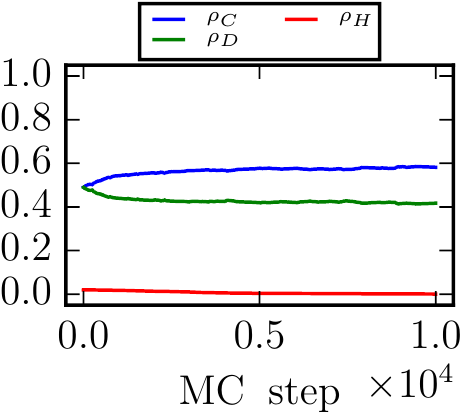}}
\subfloat[$S=0.52$, 
$T=1.48$]{\includegraphics[width=0.25\textwidth]{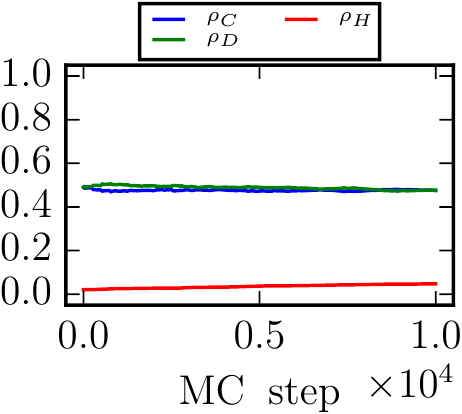}}\\
\subfloat[$S=0.48$, 
$T=1.52$]{\includegraphics[width=0.25\textwidth]{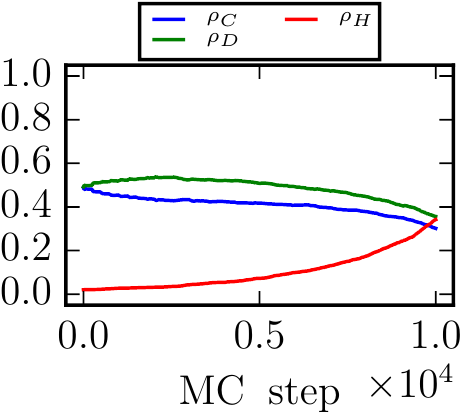}}
\subfloat[$S=0.46$, 
$T=1.54$]{\includegraphics[width=0.25\textwidth]{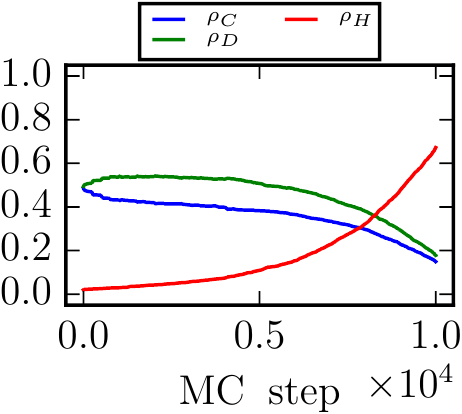}}

\caption{Full Monte Carlo history of the fractions of strategies $C$, denoted
$\rho_C$ , $D$ denoted $\rho_D$ and $H$, denoted
$\rho_H$.}\label{fig:hadamard-line-hist}
\end{figure}

Furthermore, in the snowdrift game case, the region where strategies
$C$ and $D$ coexist is much smaller compared to the classical case. This now
occurs only in the case when $S>T-1$. For other values of $S$ and $T$ in this
region we get a sharp transition to full dominance of the strategy $H$. This is
shown in more detail in Figs.~\ref{fig:hadamard-line} and
\ref{fig:hadamard-line-hist}. Studying the results across the line in the upper 
right corner of Fig.~\ref{fig:hadamard} we should note that at first we get the 
dominance of the cooperators. The fraction of cooperators starts to decrease 
with $r$ up until $r=0.5$ when we get the dominance of the miracle move. 
Analyzing the detailed histories of the strategy fractions, we note that with 
increasing $T$ the fraction of miracle moves rises faster, but the fraction of 
the strategies $C$ and $D$ is almost equal to each other.

\begin{figure}[!h]
\subfloat[$\rho_C$]{\includegraphics{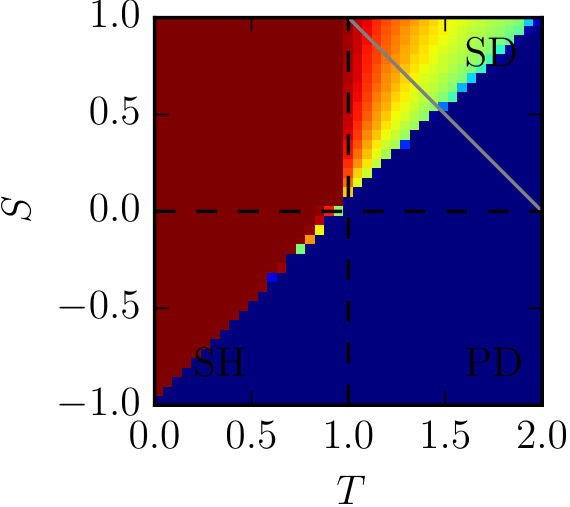}\label{fig:quantum-a}}\\
\subfloat[$\rho_D$]{\includegraphics{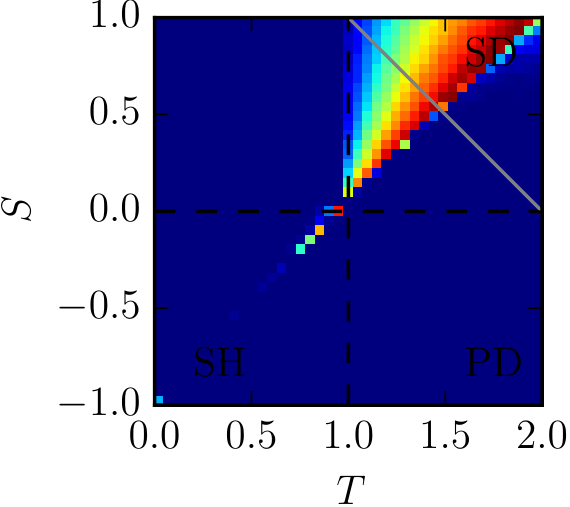}\label{fig:quantum-b}}\\
\subfloat[$\rho_Q$]{\includegraphics{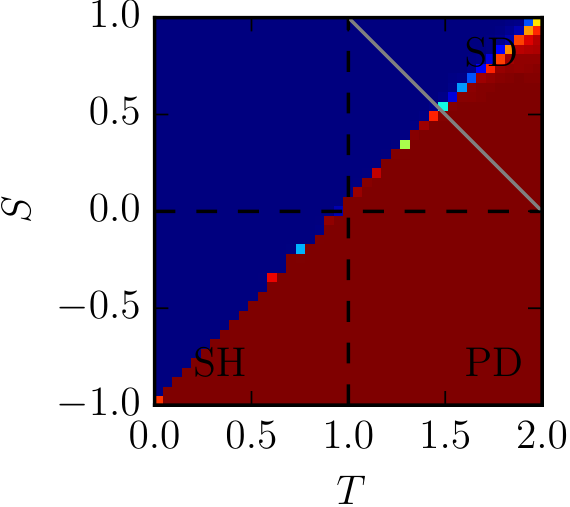}\label{fig:quantum-c}}\\
\centering\includegraphics{colorbar}
\caption{Results for the classical strategies. Panel
\protect\subref{fig:quantum-a} shows the fraction of $C$ strategy, $\rho_C$,
panel \protect\subref{fig:quantum-b} shows the fraction of strategy $D$,
$\rho_D$ and panel \protect\subref{fig:quantum-c} shows the fraction of the
strategy $Q$, $\rho_Q$. Dashed black lines mark the boundaries between the
different game types. The labels correspond to prisoner's dilemma (PD),
snowdrift (SD) and stag-hunt games (SH). The solid gray lines show the regions 
which were examined in detail.}\label{fig:quantum}
\end{figure}

\begin{figure}[!h]
\centering\includegraphics{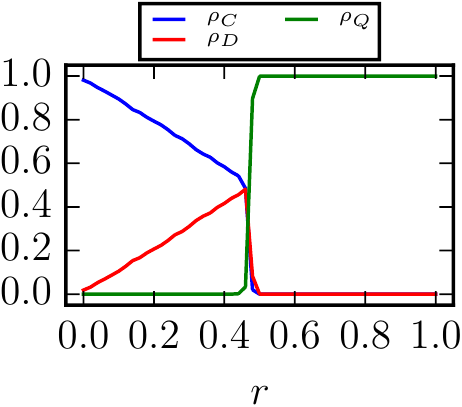}

\caption{Behavior of the fraction of strategy $C$, denoted $\rho_C$, the
fraction of $D$, $\rho_D$ and the fraction of strategy $Q$, denoted $\rho_Q$
along the line shown in the upper right corner of Fig.~\ref{fig:quantum}. The
parameter $r$ gives the values of $S=1-r$ and $T=1+r$.}\label{fig:quantum-line}
\end{figure}

\begin{figure}[!h]
\subfloat[$S=0.54$,
$T=1.46$]{\includegraphics[width=0.25\textwidth]{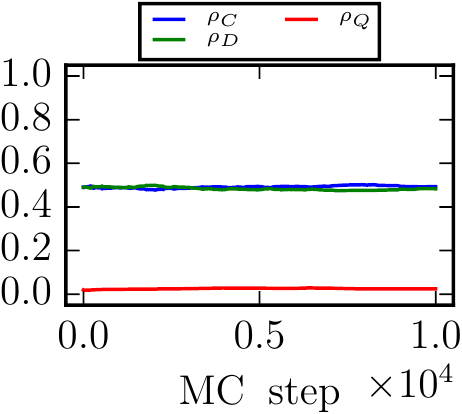}}
\subfloat[$S=0.52$,
$T=1.48$]{\includegraphics[width=0.25\textwidth]{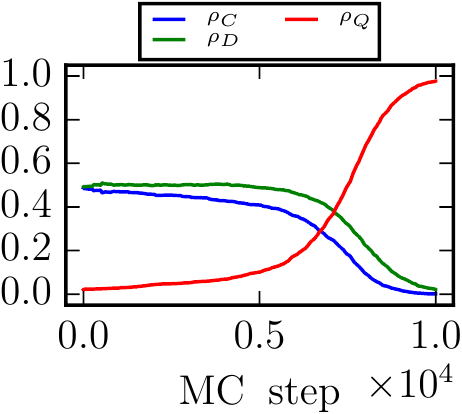}}\\
\subfloat[$S=0.5$,
$T=1.5$]{\includegraphics[width=0.25\textwidth]{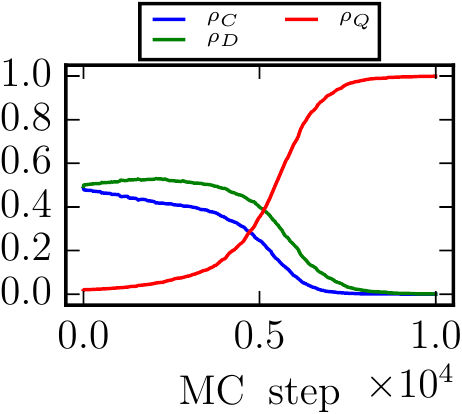}}

\caption{Full Monte Carlo history of the fractions of strategies $C$, denoted
$\rho_C$ , $D$ denoted $\rho_D$ and $Q$, denoted 
$\rho_Q$.}\label{fig:quantum-line-hist}
\end{figure}

When we introduce the strategy $Q$ with the classical ones, the behavior 
changes slightly, compared to the case studied in the case of the miracle move, 
$H$.  This is shown in Figs.~\ref{fig:quantum}, \ref{fig:quantum-line} and 
\ref{fig:quantum-line-hist}. We do not have the region in the stag-hunt game 
where there a transition between $C$ and $Q$. Instead, the entire region is 
split in half, where one is dominated by the strategy $C$ and the other by 
strategy $Q$. In the snowdrift region, the transition between coexistence of 
$C$ and $D$ and dominance of $Q$ i much sharper compared to the miracle move 
case. The prisoner's dilemma region is again dominated by the quantum strategy 
$Q$.

\begin{figure*}[!h]
\subfloat[$\rho_C$]{\includegraphics{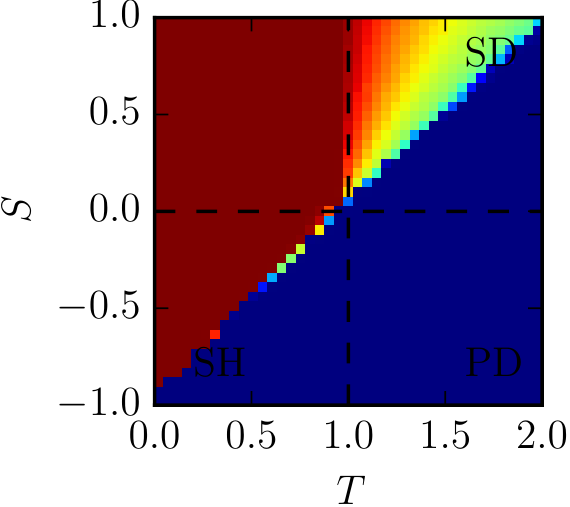}\label{fig:quantum-hadamard-a}}
 
\subfloat[$\rho_D$]{\includegraphics{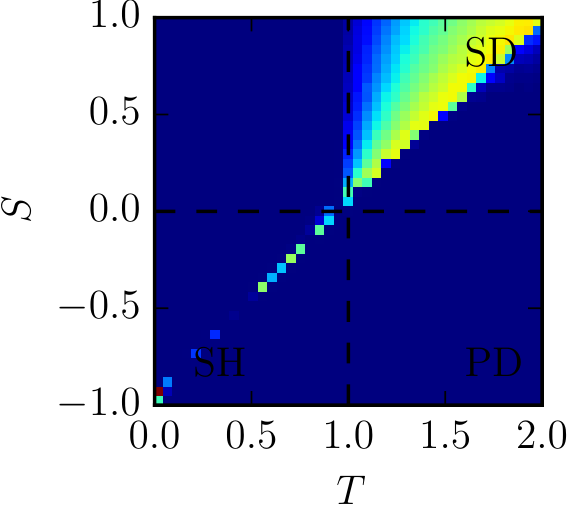}\label{fig:quantum-hadamard-b}}\\
 
\subfloat[$\rho_H$]{\includegraphics{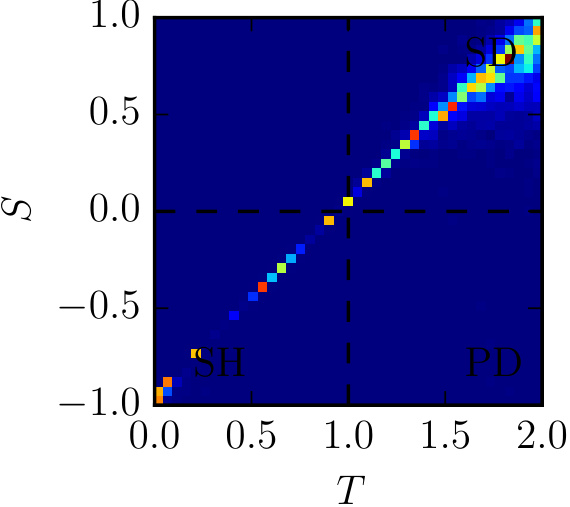}\label{fig:quantum-hadamard-c}}
 
\subfloat[$\rho_Q$]{\includegraphics{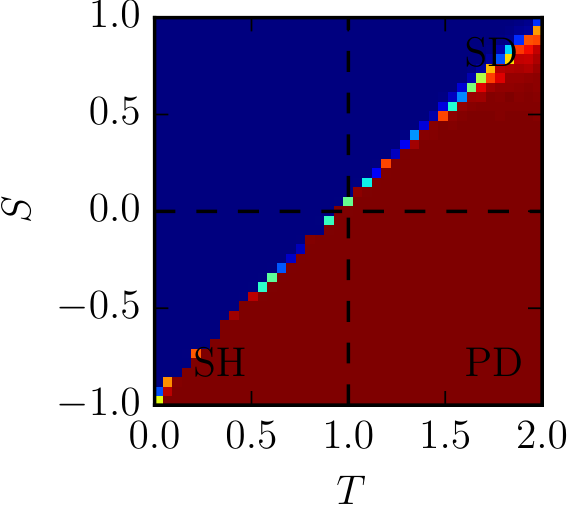}\label{fig:quantum-hadamard-d}}\\
\centering\includegraphics{colorbar}

\caption{Results for the classical strategies. Panel
\protect\subref{fig:quantum-hadamard-a} shows the fraction of $C$ strategy,
$\rho_C$, panel \protect\subref{fig:quantum-hadamard-b} shows the fraction of
strategy $D$, $\rho_D$, panel \protect\subref{fig:quantum-hadamard-c} shows the
fraction of the strategy $H$, $\rho_H$ and panel
\protect\subref{fig:quantum-hadamard-c} shows the fraction of the strategy $Q$,
$\rho_Q$. Dashed black lines mark the boundaries between the different game
types. The labels correspond to prisoner's dilemma (PD), snowdrift (SD) and
stag-hunt games (SH). The solid gray lines show the regions which were examined
in detail.}\label{fig:quantum-hadamard}
\end{figure*}

\begin{figure}[!h]
\centering\includegraphics{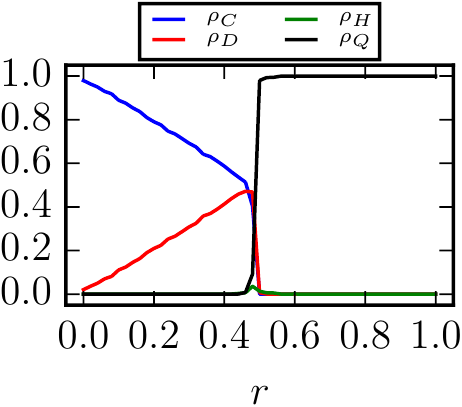}

\caption{Behavior of the fraction of strategy $C$, denoted $\rho_C$, the
fraction of $D$, $\rho_D$, fraction of strategy $H$, denoted $\rho_H$, and the
fraction of strategy $Q$, denoted $\rho_Q$ along the line shown in the upper
right corner of Fig.~\ref{fig:quantum-hadamard}. The parameter $r$ gives the
values of $S=1-r$ and $T=1+r$.}\label{fig:quantum-hadamard-line}
\end{figure}

\begin{figure}[!h]
\subfloat[$S=0.54$,
$T=1.46$]{\includegraphics[width=0.25\textwidth]{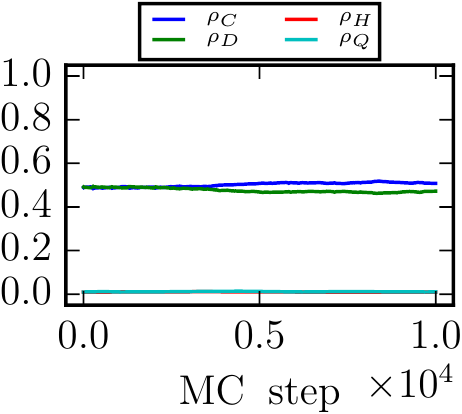}}
\subfloat[$S=0.52$,
$T=1.48$]{\includegraphics[width=0.25\textwidth]{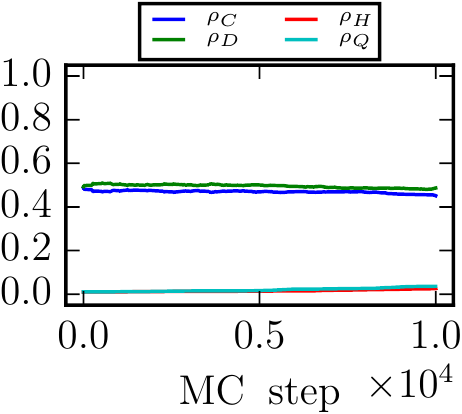}}\\
\subfloat[$S=0.5$,
$T=1.5$]{\includegraphics[width=0.25\textwidth]{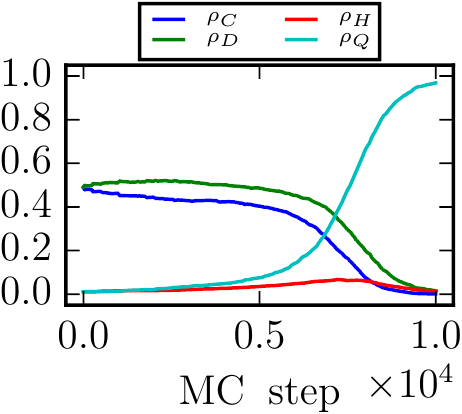}}

\caption{Full Monte Carlo history of the fractions of strategies $C$, denoted
$\rho_C$ , $D$ denoted $\rho_D$, $H$, denoted $\rho_H$, and $Q$, denoted 
$\rho_Q$.}\label{fig:quantum-hadamard-line-hist}
\end{figure}

Finally, when we introduce all four strategies, we obtain results similar to 
the case with the $Q$ strategy only. Again, all interesting regions are 
dominated by the strategy $Q$. This is shown in detail in 
Figs.~\ref{fig:quantum-hadamard}, \ref{fig:quantum-hadamard-line} and 
\ref{fig:quantum-hadamard-line-hist}.
\section{Conclusions}\label{sec:final}
To sum up, we studied evolutionary cooperation on evolving random networks with
quantum agents. We identified regions of parameters where quantum strategies
dominate the network as well as studied in detail regions with coexistence of
strategies. These are different for the classical and quantum cases.
Furthermore, introduction of the quantum miracle move introduces a new region
where the classical strategies coexists in the stag-hunt game. We should further
note that in the prisoner's dilemma region, we always get a full domination of
the quantum strategies, when they are introduced. This is consistent with
previous results~\cite{li2012quantum,pawela2013quantum}.
\section*{Acknowledgements}
Work by {\L}P was supported by the Polish National Science Centre under grant
number DEC-2011/03/D/ST6/00413.

\bibliographystyle{apsrev}
\bibliography{evolving_games}

\end{document}